\documentclass{article}
\usepackage{amsmath}
\usepackage{amsfonts}
\usepackage{amssymb}
\usepackage{amsthm}
\theoremstyle{definition}
\newtheorem{theorem}{Theorem}
\newtheorem{definition}{Definition}

\begin{document}
\title{Global Non-perturbative Deformation Quantization of a Poisson Algebra}
\author{Luther Rinehart}
\maketitle
\begin{abstract}
A proposed definition is given for the quantization of a Poisson algebra, taking the quantum product to be a geodesic on the manifold of associative products.
\end{abstract}

This paper deals with the problem of quantizing a classical algebra of observables.  Given a commutative algebra, we wish to `deform' its product to a noncommutative quantum product, thus obtaining a quantum machanical algebra of observables.  Specifically, we seek a smooth curve in the space of associative bilinear products, with initial point equal to the original commutative product.  In addition, the principles of canonical quantization suggest that the initial tangent of this curve should equal the Poisson bracket. \\
The case of the algebra of functions on a finite-dimensional symplectic vector space has long been well-understood. There, the deformed algebra is the Weyl algebra.  Its formulation as a deformation quantization was first presented by Groenewold \cite{Groenewold} and Moyal \cite{Moyal}.  Konstevich \cite{Kontsevich} has made important contributions to the case of an arbitrary Poisson manifold, including a formula for a local, perturbative expansion of the deformed product.\\
This paper proposes a global, non-perturbative characterization of the quantum product in the case of any Banach algebra with bounded Poisson bracket.  The product is defined to be a geodesic on the manifold of associative products.\\ 
\\
Let $\mathbb{A}$ be a Poisson, Banach *-algebra with 1. This means $\mathbb{A}$ is a Banach space over $\mathbb{C}$ with the following structures:\\
An associative bilinear product $\bullet:\mathbb{A}\times\mathbb{A}\rightarrow\mathbb{A}$ satisfying\\
$f\bullet(\alpha g + \beta h)=\alpha(f\bullet g)+\beta(f\bullet h)$\\
$f\bullet g = g\bullet f$\\
$f\bullet(g\bullet h)=(f\bullet g)\bullet h$\\
$\|f\bullet g\|\leq \|f\|\|g\|$\\
An identity $1\in\mathbb{A}$ satisfying\\
$1\bullet f =f$\\
$\|1\|=1$\\
An antilinear involution $*:\mathbb{A}\rightarrow\mathbb{A}$ satisfying\\
$(\alpha f + \beta g)^* = \bar{\alpha}f^* + \bar{\beta}g^*$\\
$f^{**}=f$\\
$(f\bullet g)^* = g^*\bullet f^*$\\
$\|f^*\|=\|f\|$\\
And a bilinear Poisson bracket $\mathfrak{p}:\mathbb{A}\times\mathbb{A}\rightarrow\mathbb{A}$ satisfying\\
$f\mathfrak{p}(\alpha g +\beta h)=\alpha (f\mathfrak{p}g) + \beta(f\mathfrak{p}h)$\\
$f\mathfrak{p}g=-g\mathfrak{p}f$\\
$f\mathfrak{p}(g\mathfrak{p}h)+g\mathfrak{p}(h\mathfrak{p}f)+h\mathfrak{p}(f\mathfrak{p}g)=0$\\
$f\mathfrak{p}(g\bullet h)=(f\mathfrak{p} g)\bullet h + g\bullet(f\mathfrak{p}h)$\\
$\|f\mathfrak{p}g\|\leq \|f\| \|g\|$\\

\begin{definition}
A bilinear product $\star:\mathbb{A}\times\mathbb{A}\rightarrow\mathbb{A}$ is \emph{bounded} if there exists $C>0$ such that $\|f\star g\|\leq C \|f\| \|g\| \ \forall f,g\in\mathbb{A}$.
\end{definition}
The bounded bilinear products naturally form a Banach space, denoted $BP(\mathbb{A}) $, with norm given by
\begin{equation}
\|\star\|=\sup \frac{\|f\star g\|}{\|f\| \|g\|}.
\end{equation}
\begin{definition}
A bilinear product $\star$ is a \emph{star product} if $(\mathbb{A},\star,*,1)$ is still a Banach *-algebra with 1, that is  $\forall f,g,h\in\mathbb{A}$,\\
1. $(f\star g)^* = g^*\star f^*$\\
2. $f\star 1 = 1 \star f = f$\\
3. $(f\star g)\star h = f\star(g\star h)$
\end{definition}
Let $S(\mathbb{A})$ be the space of star products.  Because the associativity condition is nonlinear, $S(\mathbb{A})$ is a submanifold of $BP(\mathbb{A})$. As a Banach space, $BP(\mathbb{A})$ has a natural derivative operator $\nabla$. Let $\tilde{\nabla}$ be the pullback of $\nabla$ to $S(\mathbb{A})$. 
\begin{definition}
The \emph{quantum product} is a smooth curve $\star(\hbar)$ in $S(\mathbb{A})$ with tangent $\dot{\star}(\hbar)$ such that $\star(0)=\bullet$, $\dot{\star}(0)=i\mathfrak{p}/2$, and which is a geodesic of $\tilde{\nabla}$:
\begin{equation}
\dot{\star}^a\tilde{\nabla}_a \dot{\star}^b =0.
\end{equation}
\end{definition}
\begin{definition}
An \emph{automorphism} of $\mathbb{A}$ is a continuous, invertible linear map $U:\mathbb{A}\rightarrow\mathbb{A}$ that preserves the structures of $\mathbb{A}$:\\
1. $U(f)\bullet U(g)=U(f\bullet g)$\\
2. $U(1)=1$\\
3. $U(f^*)=U(f)^*$\\
4. $U(f)\mathfrak{p}U(g)=U(f\mathfrak{p}g)$
\end{definition}
\begin{theorem}
The quantum product is invariant under automorphisms, that is, for any automorphism $U$,
\begin{equation}
U(f)\star U(g)=U(f\star g).
\end{equation}
\begin{proof}
Every automorphism preserves $1$ and $*$, and also preserves associativity of bilinear products.  So the submanifold $S(\mathbb{A})$ is invariant under automorphisms.  Since $U$ is linear, it preserves the derivative operator $\nabla$ and hence also the pullback $\tilde{\nabla}$. Thus it preserves geodesics on $S(\mathbb{A})$. A geodesic is uniquely determined by its initial point and initial tangent, and $U$ preserves $\bullet$ and $\mathfrak{p}$.
\end{proof}
\end{theorem}
This prescription must be altered for the case of the algebra of smooth functions on a Poisson manifold.
This space is not naturally a Banach space, so the above conditions involving the norm and boundedness do not apply.  However, $\mathbb{A}=C^\infty(M)$ does have a natural topology as a Frechet space, so the norm conditions can be replaced by the appropriate continuity conditions.  Although the space $BP(\mathbb{A})$ of continuous bilinear products is not naturally a Frechet space, one can still consider a derivative operator on it defined as usual: for $f:BP(\mathbb{A})\rightarrow BP(\mathbb{A})$ and $\star\in BP(\mathbb{A})$,
\begin{equation}
\star^a \nabla_a f \equiv \lim_{t\rightarrow 0} \frac{1}{t} \left( f(x+t\star)-f(x) \right).
\end{equation}
Then one can attempt to consider geodesics on $S(\mathbb{A})$. The topological details of this require further consideration.\\
\\
The usual quantum mechanical Weyl algebra on a phase space is infinite dimensional, so it is difficult to check whether its product is indeed a geodesic on $S(\mathbb{A})$. However, the fermionic case of the Clifford algebra is finite-dimensional, so it can be explicitly verified that the Clifford product satisfies the fermionic version of above construction, using the methods of finite-dimensional linear algebra and differential geometry.  This has been done for $\text{Cliff}(2)$ and $\text{Cliff}(3)$ (see appendix).  An easy check of whether a curve is a geodesic uses the method of Lagrange multipliers.  Suppose we are working in a vector space with inner product, and the constraint manifold is given by an equation 
\begin{equation}
f^a(x)=0,
\end{equation}
where $f$ is some function.  Then a curve $x(\hbar)$ on this manifold is a geodesic if and only if there exists a dual vector $\lambda_a(\hbar)$, the Lagrange multiplier, such that
\begin{equation}
\ddot{x}^b = \lambda_a \nabla^b f^a.
\end{equation}
In our case, the vector space is the space $BP(\mathbb{A})$ of bilinear products, and the constraint function is the associator:
\begin{equation}
{f^a}_{bde}(\star)={\star^a}_{bc}{\star^c}_{de} - {\star^c}_{bd}{\star^a}_{ce},
\end{equation}
so the geodesic equation becomes
\begin{equation}\label{geodesic}
{\ddot{\star}_i}^{\ jk}=\left( {\star^a}_{bi}{\delta^j}_d {\delta^k}_e - {\star^j}_{bd}{\delta^a}_i {\delta^k}_e \right) {\lambda_a}^{bde}.
\end{equation}
In the case of $\text{Cliff}(n)$, let $\star$ be the Clifford product. If this linear equation has a solution for $\lambda$, then the Clifford product is a geodesic.  Since $\dim\text{Cliff}(n)=2^n$ and $\dim BP(\text{Cliff}(n))=8^n$, the calculation is cumbersome for even the smallest nontrivial $n$.

\appendix
\section{Appendix}
\emph{Mathematica} code  for calculating the solution to equation \eqref{geodesic} for the Lagrange multiplier $\lambda$, in $\text{Cliff}(2)$ and $\text{Cliff}(3)$.
\begin{verbatim}
s[h_] := {{{1, 0, 0, 0}, {0, h, 0, 0}, {0, 0, h, 0}, {0, 0, 0, h^2}}, 
          {{0, 1, 0, 0}, {1, 0, 0, 0}, {0, 0, 0, -h}, {0, 0, h, 0}}, 
          {{0, 0, 1, 0}, {0, 0, 0, h}, {1, 0, 0, 0}, {0, -h, 0, 0}}, 
          {{0, 0, 0, 1}, {0, 0, 1, 0}, {0, -1, 0, 0}, {1, 0, 0, 0}}}
G := {{1, 0, 0, 0}, {0, 1, 0, 0}, {0, 0, 1, 0}, {0, 0, 0, 1}}
m[h_] := Flatten[
  Transpose[TensorProduct[s[h], G, G], {3, 4, 6, 1, 2, 5, 7}] - 
  Transpose[TensorProduct[s[h], G, G], {5, 1, 6, 4, 2, 3, 7}], 
  {{1, 2, 3}, {4, 5, 6, 7}}]
b[h_] := Flatten[s''[h]]
LinearSolve[m[h], b[h]]

s3[h_] := {
  {{1, 0, 0, 0, 0, 0, 0, 0}, {0, h, 0, 0, 0, 0, 0, 0}, 
   {0, 0, h, 0, 0, 0, 0, 0}, {0, 0, 0, h, 0, 0, 0, 0}, 
   {0, 0, 0, 0, -h^2, 0, 0, 0}, {0, 0, 0, 0, 0, -h^2, 0, 0}, 
   {0, 0, 0, 0, 0, 0, -h^2, 0}, {0, 0, 0, 0, 0, 0, 0, -h^3}}, 
  {{0, 1, 0, 0, 0, 0, 0, 0}, {1, 0, 0, 0, 0, 0, 0, 0}, 
   {0, 0, 0, 0, 0, 0, 0, 0}, {0, 0, 0, 0, 0, h, 0, 0}, 
   {0, 0, 0, 0, 0, 0, 0, -h^2}, {0, 0, 0, -h, 0, 0, 0, 0}, 
   {0, 0, 0, 0, 0, 0, 0, 0}, {0, 0, 0, 0, -h^2, 0, 0, 0}}, 
  {{0, 0, 1, 0, 0, 0, 0, 0}, {0, 0, 0, 0, 0, 0, h, 0}, 
   {1, 0, 0, 0, 0, 0, 0, 0}, {0, 0, 0, 0, 0, 0, 0, 0}, 
   {0, 0, 0, 0, 0, 0, 0, 0}, {0, 0, 0, 0, 0, 0, 0, -h^2}, 
   {0, -h, 0, 0, 0, 0, 0, 0}, {0, 0, 0, 0, 0, -h^2, 0, 0}}, 
  {{0, 0, 0, 1, 0, 0, 0, 0}, {0, 0, 0, 0, 0, 0, 0, 0}, 
   {0, 0, 0, 0, h, 0, 0, 0}, {1, 0, 0, 0, 0, 0, 0, 0}, 
   {0, 0, -h, 0, 0, 0, 0, 0}, {0, 0, 0, 0, 0, 0, 0, 0}, 
   {0, 0, 0, 0, 0, 0, 0, -h^2}, {0, 0, 0, 0, 0, 0, -h^2, 0}}, 
  {{0, 0, 0, 0, 1, 0, 0, 0}, {0, 0, 0, 0, 0, 0, 0, h}, 
   {0, 0, 0, 1, 0, 0, 0, 0}, {0, 0, -1, 0, 0, 0, 0, 0}, 
   {1, 0, 0, 0, 0, 0, 0, 0}, {0, 0, 0, 0, 0, 0, 0, 0}, 
   {0, 0, 0, 0, 0, 0, 0, 0}, {0, h, 0, 0, 0, 0, 0, 0}}, 
  {{0, 0, 0, 0, 0, 1, 0, 0}, {0, 0, 0, -1, 0, 0, 0, 0}, 
   {0, 0, 0, 0, 0, 0, 0, h}, {0, 1, 0, 0, 0, 0, 0, 0}, 
   {0, 0, 0, 0, 0, 0, 0, 0}, {1, 0, 0, 0, 0, 0, 0, 0}, 
   {0, 0, 0, 0, 0, 0, 0, 0}, {0, 0, h, 0, 0, 0, 0, 0}}, 
  {{0, 0, 0, 0, 0, 0, 1, 0}, {0, 0, 1, 0, 0, 0, 0, 0}, 
   {0, -1, 0, 0, 0, 0, 0, 0}, {0, 0, 0, 0, 0, 0, 0, h}, 
   {0, 0, 0, 0, 0, 0, 0, 0}, {0, 0, 0, 0, 0, 0, 0, 0}, 
   {1, 0, 0, 0, 0, 0, 0, 0}, {0, 0, 0, h, 0, 0, 0, 0}}, 
  {{0, 0, 0, 0, 0, 0, 0, 1}, {0, 0, 0, 0, 1, 0, 0, 0}, 
   {0, 0, 0, 0, 0, 1, 0, 0}, {0, 0, 0, 0, 0, 0, 1, 0}, 
   {0, 1, 0, 0, 0, 0, 0, 0}, {0, 0, 1, 0, 0, 0, 0, 0}, 
   {0, 0, 0, 1, 0, 0, 0, 0}, {1, 0, 0, 0, 0, 0, 0, 0}}} 
G3 := {{1, 0, 0, 0, 0, 0, 0, 0}, {0, 1, 0, 0, 0, 0, 0, 0}, 
       {0, 0, 1, 0, 0, 0, 0, 0}, {0, 0, 0, 1, 0, 0, 0, 0}, 
       {0, 0, 0, 0, 1, 0, 0, 0}, {0, 0, 0, 0, 0, 1, 0, 0}, 
       {0, 0, 0, 0, 0, 0, 1, 0}, {0, 0, 0, 0, 0, 0, 0, 1}}
m3[h_] := 
 Flatten[
 Transpose[TensorProduct[s3[h], G3, G3], {3, 4, 6, 1, 2, 5, 7}] - 
 Transpose[TensorProduct[s3[h], G3, G3], {5, 1, 6, 4, 2, 3, 7}], 
 {{1, 2, 3}, {4, 5, 6, 7}}]
b3[h_] := Flatten[s3''[h]]
LinearSolve[m3[h], b3[h]]
\end{verbatim}

\end{document}